# Gesture controlled environment using sixth sense technology and its implementation in IoT

Shubhankar Mohan, Aditi Chaudhary, Prachie Gupta, Dr. Ritu Tiwari

mohanshubhankar@gmail.com, aditichaudhary175@gmail.com, prachiegupta079@gmail.com

Indian Institute of Information Technology and Management Gwalior

## 1 ABSTRACT

This paper proposes an idea of building an interface to merge the existing technologies like Image processing, Internet of Things, Sixth sense etc. at one place to reduce the hardware restrictions imposed on a user and improve the responsiveness of the system. The wearable device comprises of a camera, a projector and its own gesture controlled environment having smart tools based on trending techniques like gesture recognition, color marker detection and speech recognition. The interface is trained using machine learning.It is also interfaced with an IoT based lab to access the lab controls remotely, enhance the security and to connect devices present in the lab.

## 2 INTRODUCTION

Sixth sense technology helps in minimizing the gap between the real and the computing world by displaying the digital information out in the real environment and allows the user to interact with it as described in [3] and [2].It requires a wearable device consisting of wide field of view camera and a projector is used to connect and communicate to things around .The device was introduced by [24]. The device brings us closer to the real environment, just like the smart class developed by [10].

Computer vision techniques have plays a major role to build up an interface for inputs acquisition and processing using image processing techniques.[18] shows how these techniques help us to derive every digital information that is in need from an image or video, which can be clubbed with other learning algorithms.Some basics and algorithm are explained in [7]. The very first way of interacting with computers using different hand gestures was first proposed by [16]. [12] , firstly adopted hand gestures to act as an interface in Human Computer Interaction.An interface based on hand gesture recognition for controlling applications like media player using computer vision techniques on a computing environment, is proposed by [25]. An artificial neural network (ANN) based gesture classifier can be used for the computing environment like [13] and [11]. Hand segmentation techniques improvises the recognition process as shown in [26].

Speech recognition, or Automatic speech recognition (ASR) is defined as the process of interpreting human speech commands in a computing environment [20]. However, Speech recognition is technically defined as the building of system or an interface for mapping voice signals to a string of words [15]. The signals are mapped to string which is then compared with the data set in order to perform the function for which the speech command was given for.

The era of IoT has greatly helped in the evolution from end-to-end communication between clients and servers on the Internet, to Internet-enabled physical devices to communicate with each other and with human. The Internet of Things (IoT) is a new generation of Internet services that enables physical devices to communicate with each other by using the World Wide Web in a wired or wireless medium. The IoT peers have to be recognized, managed and controlled to build up a connection among the devices or humans which act as the peers. They must have the ability to interact with human or other objects within the Machine to Machine (M2M) environment [27]. IoT can be implemented in a wide range of applications such as health care [6], smart cities, agriculture, smart grid, saving energy [4], home automation [22], smart building , Intelligent Traffic system [23] and more.

## 3 LITERATURE REVIEW

| Authors | Years | Remarks; |
|---------|-------|----------|
| Steve Mann | 1999 | He proposed sixth sense as a wearable device which was the combination of projector and camera. |
| P. Mistry | 2009 | He implemented Sixth sense technology to build real time applications.He used the color markers to control the system with the help of the wearable device. |
| Ala AlFuqaha | 2014 | This paper provides an overview of the Internet of Things (IoT) and gave attention on enabling technologies, protocols, and application issues. |
| Jayavardhana and Palaniswami | 2015 | This paper briefly explains Internet of Things and vital technologies which are used in daily life. |
| G.R.S. Murthy | 2010 | This paper demonstrates a hand gesture recognition in a controlled environment. |
| Piyush Kumar and Prasad | 2012 | This paper proposed hand gesture recognition technology using hand gloves and a K-NN classifier was used to detect specific gesture. |
| S.S. Rautaray | 2015 | This paper describes static hand gesture recognition using images of different shapes and orientation which are extracted from video stream in predefined environmental conditions. |

## 4 PROPOSED WORK

### 4.1 Gesture Recognition

An artificial neural network (ANN) has been used to train the interface and to be utilised as a gesture classifier for the computing environment. The software has been trained over a data set having gestures images using neural networks to achieve higher accuracy and better performance. We aim at controlling the interface using some gestures eliminating the needs of hardware.So the five phases are executed as described in the next sub sections.

*4.1.1 Hand Segmentation.* Most important task of gesture recognition is hand Segmentation (as shown in Fig: 1) based on skin colour techniques (as shown by [8, 9]). For detecting the skin color the frame captured from current video stream or in *RGB* colour space is firstly converted into *YCbCr* colour space. In *YCbCr*,

Y = luminance
Cb = blue-difference
Cr = red-difference

### (1) Conversion Method

$$Y = 0.2126 * (219/255) * R + 0.7152 * (219/255) * G + 0.0722 * (219/255) * B + 16 \qquad (1)$$

$$Cb = -0.2126/1.18556 * (224/255) * R - 0.7152/1.8556(224/255) * G + 0.05 * (219/255) * B + 128 \quad (2)$$

$$Cr = 0.5 * (224/255) * R - 0.7152/1.5748(224/255) * G + 0.0722/1.5748 * (219/255) * B + 128 \quad (3)$$

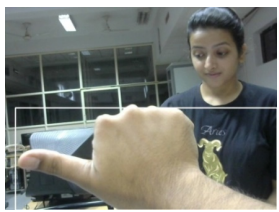

Fig. 1. Hand detection

*4.1.2 Choosing Threshold value.* The converted image is then transformed to a gray scale image to make further processing easier. By trying different threshold ranges we choose the following range to be most accurate

Cr = [130, 185]

Cb = [80, 135]

So, a gray scale image is obtained containing pixel values 0 or 255 according to

Point = 255 , if in range, 0 otherwise

*4.1.3 Feature Extraction.* Next task is to find ROI (Region of Interest)i.e we need to extract the hand.In sections where skin has been detected in frame, we detect the contour which differ in properties from their surroundings and are uniform within them. After this the largest contour is extracted re-sized to 50x50. It is then fed to Neural Net for further processing. The purpose of cropping is to increase the accuracy and decrease processing time.

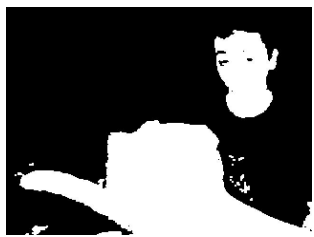

Fig. 2. Hand is extracted

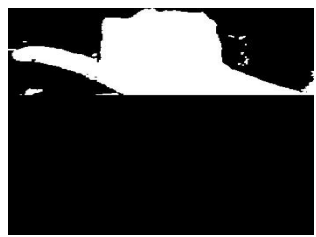

Fig. 3. Cropped hand image

*4.1.4 Training Neural Network.* After Feature extraction, we require to recognize the gesture. Here we are using Artificial Neural Networks (ANN's) as a classifier for recognizing the gesture and the performing the required action.

(1) **Training and Testing** The network was trained to classify 10 different hand gestures with a data set of 2000 pre-processed images.A similar data set with about 400 images was created as a test database and the accuracy was found out to be 100%. But the scenario changes when we talk about the real time gesture recognition. Due to light differences, sudden changes in brightness, hands sudden movement, surrounding objects in skin color range and presence of other humans the 100% accuracy can't be achieved.

(2) **Post Processing** In some papers a glove has been used to detect the gesture ([21, 28]) whereas we have used the real time environment which is effected by many factors. To overcome the error because of environmental factor we came up with an algorithm which gives accuracy up to 99.9%. It returns a gesture number or a zero, if no gesture is recognised. The algorithm is designed to reduce false positive though false negative also increases.

Table 1. Architecture of NN

| Number of Input Neurons | 2500 |
|---|---|
| Number of hidden Layers | 2 |
| Number of neurons in HL1 | 2500 |
| Number of neurons in HL2 | 1200 |
| Number of output neurons | 10 |
| Learning Rate | 0.00001 |
| Non-liner Function | Sigmoid Function |
| Epochs | 5000 |
| Bias | 1 |

## 4.2 Iot implementation in smart lab

The IoT-based Smart lab is mainly based on the integration of two technologies: Sixth sense Technology and IoT. The devices present in lab such as fans and lights are controlled through speech commands. Image processing and socket programming techniques have been used to make lab more secure via electromechanical aldrop and motion detection.

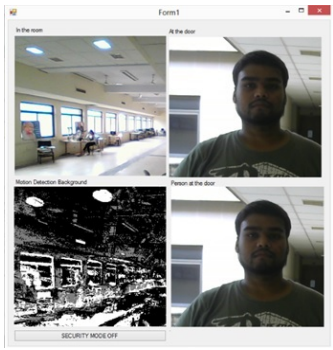

Fig. 4. IoT Lab Software

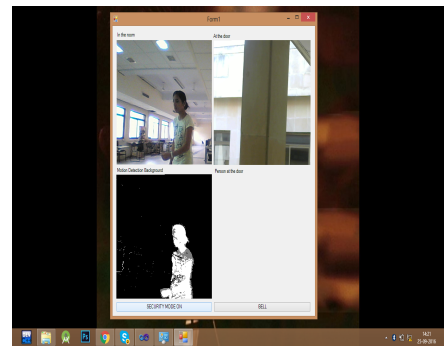

Fig. 5. When someone preaches in the room.

### 4.2.1 *Speech based controls :* Speech recognition is the basic idea which can be used to control the on/off mechanism of fans and lights regardless of user's position in the lab. In this project we have used speech recognition using Microsoft speech recognition API. In this, input data set is appointed as commands and it is loaded in grammar recognition engine. As a user gives one of the commands,

the speech command is processed and then converted to text, which is further compared with the data sets created. If it matches then micro controller takes the necessary action.

---

**ALGORITHM 1:** Processing of captured frame to recognize hand gesture

---

**Input** : Image Frame
**Output**: gen_num
*//to preprocess the input given to algorithm*
**while** *true* **do**

    *Convert RGB→ YCrCb frame*
    *Convert YCrCb → GrayscaleImage*
    *//further processing using gray scale image*
    *Detection of contours*
    *blob = largest contour detected*
    blob_cropped= *cropped(blob)*
    blob_resize= *resize(*blob_cropped*)*
    $\lfloor num, num_prob \rfloor = feed forward nn($blob_resize$)$
    *que.dequeue();*
    **if** num_prob *$>.95$* **then**
        *que.enqueue(num);*
    **else**
        *que.enqueue(0) ;*
    **end**
    **if** *que is uniform AND que[1] != 0* **then**
        ges_num = *que[1];*
    **else**
        ges_num = *0;*
    **end**
    *Action(*ges_num*);*
    *//action function deals with the response for detected gesture*
**end**

---

*4.2.2 Security mode :* In order to make lab more secure and authentic we have used the motion detection techniques to increase the security level of the lab.A wide field view camera inside the lab captures the lab environment and monitors the movement. The current frame in video stream is mainly used for computing a background model using a number of frames in fixed period of time. Further the pixel values in the current frame are compared with the current background model pixel values ([17, 19]). The deterministic method described in compares the previous and current frame by computing the local variance of the intensity ratios and the pixels count as described in [14]. This is done by using Background subtraction, motion history and pixel count.

(1) **Entrance check at door:** To increase security of lab from any unethical entrance of a person and provide an auto security check at door of lab we have used an electromechanical aldrop, its functioning is controlled via commands from micro controller. To enter the lab while the gate is close one has to press the bell, then the camera present at the door captures an image of person and sends it to lab-in-charge for authentication and permission to open the door using socket programming. An alert sound and a dialog box appear on his screen comprising the image of the person and a permission box whether to allow that person to enter or not. Depending on the permission the command is sent to the micro controller.

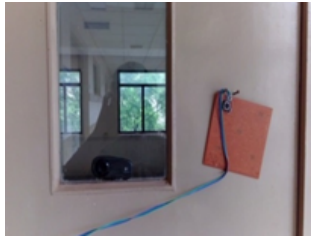

Fig. 6. Bell and camera at the door

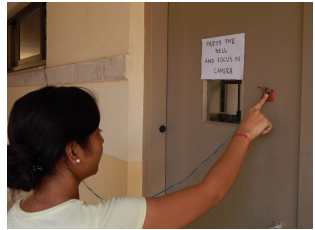

Fig. 7. Bell and camera at the door

## 4.3 Colour Marker Detection

*4.3.1 Image Segmentation.* The current frame is captured and then is converted into HSV color space from RGB color space. The HSV color frame is more ideal in extracting multiple color frames separately. In HSV, H is Hue, S is Saturation and V is value.

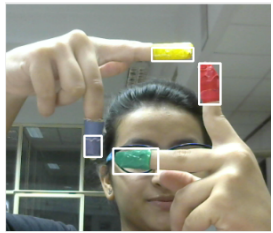

Fig. 8. Color blobs Detected

*4.3.2 Choosing Threshold value.* The converted HSV image is then again transformed into four gray scale images (one for each specified color designated to markers), to make further processing easier. We chosen threshold ranges(shown in Table:2) for four colors used in experiment. So, each of four gray scale image are obtained with containing pixel values between 0 and 255 according to

Point = 255 if in range, 0 otherwise.

Table 2. Threshold values for HSV conversion

| Color | Min Value | Max Value |
|--------|-------------|---------------|
| Red | 0, 135, 110 | 6, 255, 255 |
| Blue | 112, 53, 10 | 119, 255, 255 |
| Yellow | 68, 59, 80 | 85, 255, 255 |
| Green | 20, 165, 165 | 36, 255, 255 |

*4.3.3 Feature Extraction.* Since we are using four color, four different gray scale images (Fig:9-12 ) are the results of segmentation process, each image is then searched for ROI in the same procedure as was used to extract features in Gesture Recognition (section 2.4.1). The centroids of four blobs extracted are noted for use [1, 5].

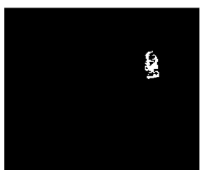
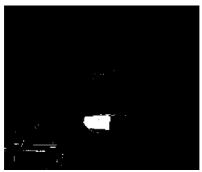
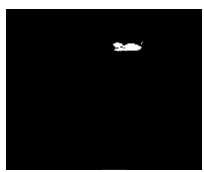
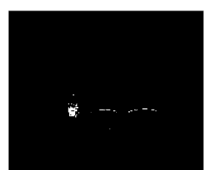

Fig. 9. Red color Detected

Fig. 10. Green color Detected

Fig. 11. Yellow color Detected

Fig. 12. Blue color Detected

## 4.4 $\iota^2$ : The Integrated Environment

Converting imagination into reality, $\iota^2$ is the integrated software controlled by the gesture and colour markers. It provides the social network interface to the user where he can access his social account without opening any browser. It can be displayed anywhere using the projector.

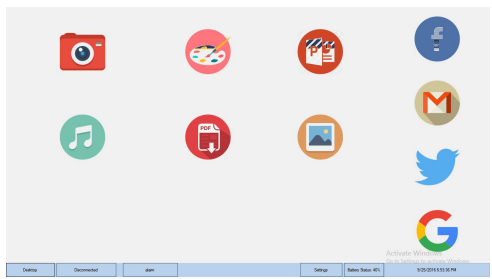

Fig. 13. $\iota^2$ software

*4.4.1 Controls of the environment.* The environment or the software is controlled using gestures and colour markers. The software is trained using the neural networks based on back propagation algorithm. Gesture classifier is used to recognize the gesture and perform the necessary actions whereas colour marker based mouse controller is used for the primary pointer operation.

**Gesture Controls.** Various gestures are used to control the environment. Basic function of an environment consisting of different applications is to open or close an application, minimising or maximising the application etc. Basically instead of doing all these things manually we use gestures to automate the environment. This is implemented using a ANN gesture classifier trained using the gesture data sets. When a user shows a gesture in front of camera the gesture is detected and then fed to classifier which in turn recognises the gesture and perform the required action.

**Color Based mouse controller.** Having a pointer hardware device also restricts a user in many ways . This can be avoided by using a color based markers for controlling the pointer in the environment for pointer operations. Different colors are detected using the blob detection techniques along with the thresholds for different colors. The basic controls are right click, left click, scrolling and dragging. The movement of mouse cursor is calculated with respect to centroids of color markers through following equations.
movement(X',Y') can be represented as:

$$X' = \frac{(\alpha * centroid\_x * screen\_size\_x)}{video\_width} + \delta \tag{4}$$

$$Y' = \frac{(\beta * centroid\_y * screen\_size\_y)}{video\_height} + \gamma \tag{5}$$

where $\alpha$, $\beta$ are positive real numbers, $\gamma$, $\delta$ are integers , (centroid_x, centroid_y) is coordinate of centroid green marker and (screen_size_x, screen_size_y) is size of primary screen.

*4.4.2* **Gesture and colour marker based tools.** The integrated environment is equipped with gesture, speech and marker based tools which makes the life of a user more convenient and easy. These tools require less hardware and give a lot more in return. It is handy for the personal as well as professional purposes. The tools are their working is described below.

(1) G-Presenter
(2) Marker based Paint
(3) Speech based music player
(4) Click less-Camera
(5) Auto-Scroll PDF
(6) G-Photo View
(7) Social network panel

## 5 CONCLUSION AND FUTURE SCOPE

Earlier work in this field was very segregated, different technologies though exists in one form or another but were available for use on single platform or independently. This combination of technologies can be used to build various interface one of them is gesture controlled environment using machine learning and its implementation in Internet of Things.For sixth sense technology integration of gesture recognition and color marker reduces the error as some main executing commands are executed via combination of both which reduces the chances of failure. We overlap the virtual environment with the real environment.The environment can be controlled using hand gesture and color markers.IoT can be used to further integrate smart objects to the envirinment like the IoT lab mentioned above.

**Future Scope:**
    The work can be further implemented in various other field like education, medical, defense and many more. The virtual and real environment are converging so this project may help in reducing the gap.